%% file: QCM.tex
\input{LocalHeader}\usepackage{verbatim}

\newcommand\Ix{{\mathsf{Ix}}}
\newcommand\MajIx{{\mathsf{MajIx}}}
\newcommand\Tr{{\mathsf{Tr}}}
\date{}
\newident{\EQ}{EQ}
\newident{\NE}{NE}

\title{Two Results about Quantum Messages}
\author{Hartmut Klauck%
  \thanks{Division of Mathematical
Sciences, Nanyang Technological University, Singapore 637371 \& Centre for Quantum Technologies,
National University of Singapore, Singapore 117543. \hbox{E-mail}:~{\tt hklauck@gmail.com}.
This work is funded by the Singapore Ministry of Education (partly through the Academic Research Fund Tier 3 MOE2012-T3-1-009)
and by the Singapore National Research Foundation.}
\and{Supartha Podder}\thanks{Centre for Quantum Technologies. \hbox{E-mail}:~{\tt supartha@gmail.com}.}
}

\begin{document}

\maketitle

\thispagestyle{empty}

\abstr{

We show two results about the relationship between quantum and classical messages. Our first contribution is to show how to replace a quantum message in a one-way communication protocol by a deterministic message, establishing that for all partial Boolean functions $f:\{0,1\}^n\times\{0,1\}^m\to\{0,1\}$ we have $D^{A\to B}(f)\leq O(Q^{A\to B,*}(f)\cdot m)$. This bound was previously known for total functions, while for partial functions this improves on results by Aaronson \cite{aaronson:advicecommj,aaronson:qlearnability}, in which either a log-factor on the right hand is present, or the left hand side is $R^{A\to B}(f)$, and in which also no entanglement is allowed.

In our second contribution we  investigate the power of quantum proofs over classical proofs. We give the first example of a scenario, where quantum proofs lead to exponential savings in computing a Boolean function. The previously only known separation between the power of quantum and classical proofs is in a setting where the input is also quantum \cite{aaronson:qproof}.

We exhibit a partial Boolean function $f$, such that there is a one-way quantum communication protocol receiving a quantum proof (i.e., a protocol of type QMA) that has cost $O(\log n)$ for $f$, whereas
every one-way quantum  protocol for $f$ receiving a classical proof (protocol of type QCMA) requires communication $\Omega(\sqrt n/\log n)$.
}

\newpage

\setcounter{page}{1}

\sect[intro]{Introduction}

The power of using quantum messages over classical messages is a central topic in information and communication theory. It is always good to understand such questions well in the simplest settings where they arise. An example is the setting of one-way communication complexity, which is rich enough to lead to many interesting results, yet accessible enough for us to show results about deep questions like the relationship between different computational modes, e.g.~quantum versus classical or nondeterministic versus deterministic.

\subsection{One-way Communication Complexity}
Perhaps the simplest question one can ask about the power of quantum messages is the relationship between quantum and classical one-way protocols. Alice sends a message to Bob in order to compute the value of a function $f:\{0,1\}^n\times\{0,1\}^m\to\{0,1\}$. Essentially, Alice communicates a quantum state and Bob performs a measurement, both depending on their respective inputs. Though deceptively simple, this scenario is not at all fully understood. Let us  just mention the following open problem: what is the largest complexity gap between quantum and classical protocols of this kind for computing a total Boolean function? The largest gap known is a factor of 2, as shown by Winter \cite{winter:equality}, but for all we know there could be examples where the gap is exponential, as it indeed is for certain partial functions \cite{gavinsky:one-way}.

An interesting bound on such speedups can be found by investigating the effect of replacing quantum by classical messages.
Let us sketch the proof of such a result. Suppose a total Boolean function $f$ has a quantum one-way protocol with communication $c$, namely Alice sends $c$ qubits to Bob, who can decide $f$ with error 1/3 by measuring Alice's message. We allow Alice and Bob to share an arbitrary input-independent entangled state.
Extending Nayak's random access code bound \cite{nayak:qfa} Klauck \cite{klauck:qpcom} showed that $Q^{A\to B,*}(f)\geq \Omega(VC(f))$, where $Q^{A\to B,*}(f)$ denotes the entanglement-assisted quantum one-way complexity of $f$, and $VC(f)$ the Vapnik-Chervonenkis dimension of the communication matrix of $f$. Together with Sauer's Lemma \cite{sauer:lemma} this implies that $D^{A\to B}(f)\leq O(Q^{A\to B, *}(f)\cdot m)$, where $m$ is the length of Bob's input. See also \cite{JZ09} for a related result.

A result such as the above is much more interesting in the case of partial functions. The reason is that for total functions a slightly weaker statement follows from a weak version of the random access code bound, which can be (and indeed has been \cite{ambainis:rac}) established by the following argument: boost the quantum protocol for $f$ until the error is below $2^{-2m}$, where $m$ is Bob's input length. Measure the message sent by Alice with {\em all} the measurements corresponding to Bob's inputs (this can be done with small total error) in order to determine Alice's row of the communication matrix and hence her input. This is a hard task by standard information theory facts (Holevo's bound). When considering partial functions a disaster happens: Bob does not know for which of his inputs $y$ the value $f(x,y)$ is defined. If Bob measures the message for $x$ with the observable for $y$ and $f(x,y)$ is undefined any acceptance probability is possible and the message state can be destroyed.

Aaronson \cite{aaronson:advicecommj} circumvented this problem in the following way: Bob now tries to learn Alice's message. He starts with a guess (the totally mixed state) and keeps a classical description of his guess. Alice also always knows what Bob's guess is. Bob can simulate quantum measurements by brute-force calculation: for any measurement operator Bob can simply calculate the result from his classical description. Alice can do the same. Since Bob has some $2^m$ measurements he is possibly interested in, Alice can just tell him on which of these he will be wrong. Bob can then adjust his quantum state accordingly, and Aaronson's main argument is that he does not have to do this too often before he reaches an approximation of the message state. Note that Bob might never learn the message state if it so happens that all measurements are approximately correct on his guess. But if he makes a certain number of adjustments he will learn the message state and no further adjustments are needed.

Let us state Aaronson's result from \cite{aaronson:advicecommj}.

\begin{fact}
$D^{A\to B}(f)\leq O(Q^{A\to B}(f)\cdot\log(Q^{A\to B}(f))\cdot m)$ for all partial Boolean $f:\{0,1\}^n\times\{0,1\}^m\to\{0,1\}$.
\end{fact}

Aaronson later proved the following result, that removes the log-factor at the expense of having randomized complexity on the left hand side.

\begin{fact}
$R^{A\to B}(f)\leq O(Q^{A\to B}(f)\cdot m)$ for all partial Boolean $f:\{0,1\}^n\times\{0,1\}^m\to\{0,1\}$.
\end{fact}

Our first result is the following improvement.

\begin{result}
$D^{A\to B}(f)\leq O(Q^{A\to B,*}(f)\cdot m)$ for all partial Boolean $f:\{0,1\}^n\times\{0,1\}^m\to\{0,1\}$.
\end{result}

Hence we remove the log-factor, and we allow the quantum communication complexity on the right hand side to feature prior entanglement between Alice and Bob. Arguably, looking into the entanglement assisted case (which is interesting for our second main result) led us to consider a more systematic progress measure than in Aaronson's proof, which in turn allowed us to analyze a different update rule for Bob that also works for protocols with error 1/3, instead of extremely small error as used in \cite{aaronson:advicecommj}, which is the cause of the lost log-factor.

We note that this result can be used to slightly improve on the ``quantum-classical'' simultaneous message passing lower bound for the Equality function by Gavinsky et al.~\cite{GRW08_Si_Co}, establishing a tight $\Omega(\sqrt n)$ lower bound on the complexity of Equality in a model where quantum Alice and classical Bob (who do not share a public coin or entanglement) each send messages to the referee. The tight lower bound has also recently been established via a completely different and simpler method \cite{GK14} (as well as generalized to a nondeterministic setting). Our result (as well as the one in \cite{GK14}) allows a generalization to a slightly stronger model: Alice and the referee may share entanglement.

\subsection{The Power of Quantum Proofs}

We now turn to the second result of our paper, which is philosophically the more interesting. Interactive proof systems are a fundamental concept in computer science. Quantum proofs have a number of disadvantages:
reading them may destroy them, errors may occur during verification, verification needs some sort of quantum machine, and it may be much harder to provide them than classical proofs. The main hope is that quantum proofs can in some situations be verified using fewer resources than classical proofs. Until now such a hope has not been verified formally. In the fully interactive setting Jain et al.~have shown that the set of languages recognizable in polynomial time with the help of a quantum prover is equal to the set where the prover and verifier are classical (i.e., IP=QIP \cite{qip=pspace}).

The question remains open in the noninteractive setting. A question first asked by Aharonov and Naveh \cite{aharonov:qnp} and meriting much attention, is whether proofs that are quantum states can ever be easier to verify than classical proofs (by quantum machines) in the absence of interaction, i.e., whether the class QMA is larger than its analogue with classical proofs but quantum verifiers, known as QCMA.
An indication that quantum proofs may be powerful was given by Watrous \cite{watrous:group}, who described an efficient QMA black box algorithm for deciding nonmembership in a subgroup. However, Aaronson and Kuperberg \cite{aaronson:qproof} later showed how to solve the same problem efficiently using a classical witness, giving a QCMA black box algorithm for the problem. They also introduced a quantum problem, for which they show that QMA black box algorithms are more efficient than QCMA black box algorithms. Using a quantum problem to show hardness for algorithms using classical proofs seems unfair though and a similar separation has remained open for Boolean problems.

In our second main result we compare the two modes of noninteractive proofs and quantum verification for a Boolean function in the setting of one-way communication complexity. More precisely we exhibit a partial Boolean function $f$, such that the following holds. $f$ can be computed in a protocol where a prover who knows $x,y$ can provide a quantum proof to Alice, and Alice sends quantum message to Bob, such that the total message length (proof plus message Alice to Bob) is $O(\log n)$. In the setting where a prover Merlin (still knowing all inputs) sends a classical proof to Alice, who sends a quantum message to Bob, the total communication is $\Omega(\sqrt n/\log n)$.

\begin{result}
There is a partial Boolean function $f$ such that $QMA^{A\to B}(f)=O(\log n)$, while $QCMA^{A\to B,*}(f)=\Omega(\sqrt n/\log n )$.
\end{result}
We note that this is the first known exponential gap between computing Boolean functions in a QCMA and a QMA mode in any model of computation. Also, the lower bound is not too far from being tight, since there is an obvious upper bound of $O(\sqrt n\log n)$ for the problem.

So where does the power of quantum proofs come from in our result? Raz and Shpilka \cite{raz:qproof} show that QMA one-way protocols are as powerful as QMA two-way protocols. Their proof uses a quantum witness that is a superposition over the messages of different rounds.
We show that for a certain problem with an efficient QMA protocol there is no efficient one-way QCMA protocol. Hence the weakness of classical proofs here is the impossibility of compressing interaction as in the QMA case.

\sect[prelim]{Preliminaries}

\subsection{Quantum}

For basic quantum background we refer to \cite{nielsen&chuang:qc}.

\subsection{Communication Complexity Models}

We assume familiarity with communication complexity,
referring to~\cite{kushilevitz&nisan:cc} for more details about classical communication complexity
and~\cite{wolf:qccsurvey} for quantum communication complexity.

For a partial Boolean function $f:\{0,1\}^n\times\{0,1\}^m\times\{0,1,\bot\}$, where $\bot$ stands for ``undefined'' the communication matrix $A_f$ has rows labeled by $x\in\{0,1\}^n$ and columns labeled by $y\in\{0,1\}^m$, and entries $f(x,y)$.
A protocol for $f$ is correct, if it gives the correct output for all $x,y$ with $f(x,y)\neq \bot$ (with certainty for deterministic protocols, and with probability 2/3 for quantum protocols). A protocol is one-way, if Alice sends a message to Bob, who computes the function value, or vice versa.
We denote by $D^{A\to B}(f)$ the deterministic one-way communication complexity of a function $f$, when Alice sends the message to Bob.

Two rows $x,x'$ of $A_f$ are {\em distinct}, if there is a column $y$, such that $f(x,y)=1$ and $f(x',y)=0$ or vice versa, i.e., the function values differ on some defined input. Note that being not distinct is not an equivalence relation: $x,x'$ can be not distinct, as well as $x,x''$, while $x,x'$ are distinct. Nevertheless a one-way protocol for $f$ needs to group inputs $x$ into messages such that no two distinct $x,x'$ share the same message.\footnote{It is instructive to consider the function $f(x,i;y,j)=x_{i\oplus j}$ under the promise that $x=y$. This function has only $n$ {\em distinct} rows and columns, and $D^{A\to B}(f)=D^{B\to A}(f)\leq O(\log n)$. Nevertheless $A_f$ has many more actual rows and columns. Trying to reduce the number of actual columns to a set of distinct columns increases the number of distinct rows. Hence one has to be careful when considering partial functions.}

Similar to the above, we denote by $Q^{A\to B}(f)$ the quantum one-way communication complexity of $f$ with error 1/3. This notion is of course asymptotically robust when it comes to changing the error to any other constant. $Q^{A\to B,*}(f)$ denotes the complexity if Alice and Bob share entanglement.

We now define some more esoteric modes of communication that extend the standard nondeterministic mode to the quantum case.
We restrict our attention to one-way protocols.
\begin{definition}
In a one-way MA-protocol there are 3 players Merlin, Alice, Bob. Merlin sends a classical message to Alice, who sends a classical message to Bob, who gives the output.
Alice and Bob share a public coin, which is not seen by Merlin.
For a Boolean function $f:\{0,1\}^n\times\{0,1\}^m\to\{0,1\}$ the protocol is correct, if for all 1-inputs there is a message from Merlin, such that with probability 2/3 Bob will accept, whereas for all 0-inputs, and all messages from Merlin, Bob will reject with probability 2/3.
The communication complexity is defined as usual and denoted by $MA^{A\to B}(f)$.

A one-way QCMA-protocol is defined similarly, but whereas Merlin's message is still classical, Alice can send a quantum message to Bob, and Alice and Bob may share entanglement. The complexity with shared entanglement is denoted $QCMA^{A\to B,*}(f)$.

In a one-way QMA-protocol also Merlin's message may be quantum. The complexity is denoted by $QMA^{A\to B}(f)$ in the case where no entanglement is allowed.

\end{definition}

\subsection{Quantum Information Measures}

In this paper we need only a few well established notions of information and distinguishability.
A density matrix is a positive semidefinite matrix of trace 1. Density matrices will also be referred to as quantum states in this paper.

\begin{definition}
The von Neumann entropy of a quantum state $\rho$ is $S(\rho)=-\Tr\rho\log\rho$.

The relative von Neumann entropy of quantum states $\rho,\sigma$ is $S(\rho||\sigma)=\Tr\rho\log\rho-\Tr\rho\log\sigma$ if $\mbox{supp }\rho\subseteq\mbox{supp }\sigma$, otherwise $S(\rho||\sigma)=\infty$.

The relative min-entropy of $\rho,\sigma$ is $S_{\infty}(\rho||\sigma)=\inf\{c:\sigma-\rho/2^c \mbox{ is positive semidefinite}\}$.
\end{definition}

It is easy to see that $S(\rho||\sigma)\leq S_{\infty}(\rho||\sigma)$, see \cite{datta:minentropy} for a proof.
An important measure of how far apart quantum  states are is the trace distance.

\begin{definition}
The trace norm of a Hermitian operator $\rho$ is defined as $||\rho||_{t}=\Tr\sqrt{\rho\rho^\dagger}$.

The trace distance between $\rho$ and $\sigma$ is $||\rho-\sigma||_{t}$.
\end{definition}

We list two well known facts. First, Uhlmann monotonicity.
\begin{fact}
If $\tilde{\rho},\tilde{\sigma}$ result from measuring $\rho,\sigma$ then $S(\tilde{\rho}||\tilde{\sigma})\leq S(\rho||\sigma)$.
\end{fact}

Secondly, the quantum Pinsker inequality \cite{qpinsker}, see also \cite{kntz:interaction}.

\begin{fact}
$||\rho-\sigma||_t\leq \sqrt{2\ln2\ S(\rho||\sigma)}$.
\end{fact}

We note that any two states that are close in trace distance are hard to distinguish by any measurement, namely the classical distance between the measurement results is at most the trace distance of the measured states.

\section{Making Quantum Messages Deterministic}

\begin{theorem}\label{thm1}
For every partial Boolean function $f:\{0,1\}^n\times\{0,1\}^m\to\{0,1\}$ we have
 $D^{A\to B}(f)\leq O(Q^{A\to B,*}(f)\cdot m)$.
 \end{theorem}

We note that in the case of total functions the theorem follows from a result in \cite{klauck:qpcom} combined with Sauer's lemma \cite{sauer:lemma}, and that two weaker versions of the theorem have been proved by Aaronson: in \cite{aaronson:advicecommj} he shows the result with an additional log-factor on the right hand side, and without allowing entanglement, in \cite{aaronson:qlearnability} with $R^{A\to B}(f)$  on the left hand side (and no additional log-factor), but again without entanglement.

Our proof follows Aaronson's main approach in \cite{aaronson:advicecommj}, in which Bob maintains a classical description of a quantum state as his guess for the message he should have received, and Alice informs him about inputs on which this state will perform badly, so that he can adjust his guess. His goal is to either get all measurement results approximately right, or to learn the message state. We will refer to these states as the current guess state, and the target state.

We deviate from Aaronson's proof in two ways. First, we work with a different progress measure that is more transparent than Aaronson's, namely the relative entropy between the target state and the current guess. This already allows us to work in the entanglement-assisted case.

Secondly, we modify the rule by which Bob updates his guess. In Aaronson's proof Bob projects his guess state onto the subspace on which the target state has a large projection (because the message is accepted by the corresponding measurement with high probability). This has the drawback that one cannot use the actual message state of the protocol as the target state, because that state usually has considerable projection onto the orthogonal complement of the subspace, making the relative entropy infinitely large! Hence Aaronson uses a boosted and projected message state as the target state. This state is close to the actual message state thanks to the boosting, and projection of the guess state now properly decreases the relative entropy, since the target state is fully inside the subspaces. The boosting step costs exactly the log-factor we aim to remove.

So in the situation where Bob wants to update his guess state $\sigma$, knowing that the target state $\rho$ will be accepted with probability $1-\epsilon$ when measuring the observable consisting of subspace $V_y$ and its complement, we let Bob replace $\sigma$ with the mixture of $1-\epsilon$ times the projection onto $V_y$ and $\epsilon$ times the projection onto $V_y^\perp$. The main part of the proof is then to show that this decreases the relative entropy $S(\rho||\sigma)$ given that $\Tr(V_y\sigma)<1-10\sqrt\epsilon$, i.e., in case $\sigma$ was not good enough already. Eventually either all measurements can be done by Bob giving the correct result, or the current guess state $\sigma$ satisfies $S(\rho||\sigma)\leq 5\sqrt\epsilon$, in which case $\rho$ and $\sigma$ are also close in the trace distance meaning that any future measurement will give almost the same results on both states.

\begin{proof}
Fix any entanglement assisted one-way protocol with quantum communication $q=Q^{A\to B,*}(f)$.
Using standard boosting we may assume that the error of the protocol is at most $\epsilon=10^{-6}$ for any input $x,y$. This increases the communication by a small constant factor at most.

Using teleportation we can replace the quantum communication by $2q$ classical bits of communication at the expense of adding $q$ EPR-pairs to the shared entangled state. Let $|\phi\rangle$ denote the entangled state shared by the new protocol. We can assume this is a pure state, because if this is not the case we may consider any purification, and Alice and Bob can ignore the purification part. Note that we do not restrict the number of qubits used in $|\phi\rangle$.

In the protocol, for a given input $x$ Alice has to perform a unitary transformation on her part of $|\phi\rangle$ (we assume that any extra space used is also included in $|\phi\rangle$ and that measurements are replaced by unitaries) and then sends a classical message. Bob first applies the unitary from the teleportation protocol (which only depends on the classical message).
Let's denote the state shared by Alice and Bob at this point by $|\phi_x\rangle$.
 Following this Bob performs a measurement (depending on his input $y$) on his part of $|\phi_x\rangle$. This measurement determines the output of the protocol on $x,y$. We may assume by standard techniques that Bob's measurements are projection measurements, and that the subspaces used in the projection measurements have dimension $d/2$, where $d$ is the dimension of the underlying Hilbert space.

Recall that $|\phi\rangle$ and $|\phi_x\rangle$ are bipartite states shared by Alice and Bob.
 Let $\rho=\Tr_A|\phi_x\rangle\langle\phi_x|$ and $\sigma_1=\Tr_A|\phi\rangle\langle\phi|$, i.e., the states when Alice's part is traced out.  Bob wants to learn $\rho$ in order to be able to determine all measurement results on $\rho$. We show how to do this with $O(m\cdot q)$ bits of deterministic communication from Alice.
Note that the state $\sigma_1$ is known to Bob in the sense that he knows its classical description.

Since Alice's local operations do not change Bob's part of $|\phi\rangle\langle\phi|$, the difference between $\rho$ and $\sigma_1$ is introduced via the correction operations in the teleportation protocol that Bob applies after he receives Alice's message. But with probability $2^{-2q}$ Bob does not have to do anything, i.e., when Alice's message is the all 0-s string. This implies that \[\sigma_1=\frac{1}{2^{2q}}\rho+\theta,\]
for some positive semidefinite $\theta$ with trace $1-1/2^{2q}$.
Hence we get that \[S(\rho||\sigma_1)\leq S_{\infty}(\rho||\sigma_1)\leq 2q.\]
In other words, Bobs target $\rho$ and initial guess $\sigma_1$ have small relative entropy.

We can now describe the protocol. Bob starts with the classical description of $\sigma_1$. This state is also known by Alice, since it does not depend on the input. Throughout the protocol Bob will hold states $\sigma_i$, which will be updated when needed, using information provided by Alice.
Bob also has a set of measurement operators $P_y, I-P_y$ for all his inputs $y$.
Bob and Alice each loop over his inputs $y$, and compute $p_y=\Tr(P_y \sigma_i)$. This is the acceptance probability, if $\sigma_i$ is measured with the measurement for his input $y$. Alice also computes  $p'_y=\Tr(P_y\rho)$ which is the acceptance probability of the quantum protocol. If $p_y$ and $p'_y$ are too far apart, Alice will notify Bob of the correct acceptance probability on $y$ (with precision $\epsilon^2$), which takes $m+O(1)$ bits of communication.

Alice does not send a message if $f(x,y)$ is undefined, because the acceptance probability on such inputs is irrelevant. Suppose $p'_y=1-\epsilon_y$, where $\epsilon_y\leq\epsilon$ is the error on $x,y$ (and $f(x,y)=1$), but $p_y=1-a$ for some $1>a\geq10\sqrt\epsilon$. If this is not the case the measurement for $y$ applied to $\sigma_i$ already yields the correct result and no information from Alice is needed. So if $p_y,p'_y$ are far apart Alice will send $y$ (using $m$ bits) and $\epsilon_y$ as a floating point number with precision $\epsilon^2$ (using $O(1)$ bits).

Bob then adjusts $\sigma_i$ to obtain a state $\sigma_{i+1}$. Suppose he knows the correct $\epsilon_y$ (the difference between $\epsilon_y$ and its approximation sent by Alice will be irrelevant). This means that $\Tr(P_y\sigma_i)=1-a$ but $\Tr(P_y\rho)=1-\epsilon_y$. $P_y$ is the projector onto a subspace $V_y$. We have assumed that each $V_y$ has dimension $d/2$ if $d$ is the dimension of the underlying Hilbert space. Let $B_i$ denote an orthonormal basis, in which the first $d/2$ elements span $V_y$, and the remaining $d/2$ span $V_y^\perp$. Furthermore in this basis the upper left and lower right quadrants of $\sigma_i$ are diagonal.
Hence $\sigma_i$ will look like this:
$$
\sigma_i=\left(
\begin{array}{rr}
A &B\\
B^* &D\\
\end{array}
\right)=
\left(
\begin{array}{rcrrcr}
    * &    0 &    0 &    * &    \cdots &    *\\
    0 &    \ddots &    0 &   * &   \vdots &    *\\
   0 &    0 &    * &    * &   \cdots &    *\\
   * &    \cdots &    * &    * &    0 &    0\\
    * &    \vdots &    * &    0 &    \ddots &    0\\
    * &    \cdots &    * &    0 &    0&    *
\end{array}
\right),
$$

where $\Tr(A)=1-a$ and $\Tr(D)=a$.
We can now define

$$
\sigma_{i+1}=\left(
\begin{array}{cc}
\frac{1-\epsilon_y}{1-a}A\hspace{0cm}& 0\\
0& \hspace{0cm}\frac{\epsilon_y}{a}D\\
\end{array}
\right),
$$

$\sigma_{i+1}$ is diagonal in our basis $B_i$.
Clearly $\sigma_{i+1}$ would perform exactly as desired on measurement $P_y, I-P_y$. But Bob already knows the function value on $y$ and can carry on with the next $y$.

Before we continue we have to argue that the case $a=1$ can never happen. Since $\Tr(P_y\rho)=1-\epsilon_y>0$ the state $\rho$ has a nonzero projection onto $V_y$. If $a=1$ then $\sigma_i$ sits entirely in $V_y^\perp$, and hence $S(\rho||\sigma_i)=\infty$. But since we start with a finite $S(\rho||\sigma_1)$ and only decrease that value the situation $a=1$ is impossible.

Coming back to the protocol, it is obvious that Bob will learn the correct value of $f(x,y)$ for all $y$ such that $f(x,y)$ is defined. Hence the protocol is deterministic and correct. The remaining question is how many times Alice has to send a message to Bob.
We will show that this happens at most $O(Q^{A\to B,*}(f)/\sqrt\epsilon)$ times, which establishes our theorem.

The main claim that remains to be shown is the following.

\begin{claim}
$S(\rho||\sigma_i)\geq S(\rho||\sigma_{i+1})+a/2$ if $a\geq 10\sqrt\epsilon$.
\end{claim}

This establishes the upper bound on the number of messages, because the relative entropy, which starts at $2q$ is decreased by $a/2\geq5\sqrt\epsilon$ for each message. After at most $2q/(5\sqrt\epsilon)$ iterations the protocol has either ended (in which case Bob might never learn $\rho$, but will still know all measurement results), or we have
\[S(\rho||\sigma_T)\leq5\sqrt\epsilon.\]

To see this assume we are still in the situation of the claim. The claim states that the relative entropy can be reduced by $a/2$ as long as $a\geq 10\sqrt\epsilon$. So the process stops (assuming we don't run out of suitable $y's$)  no earlier than when $S(\rho||\sigma_i)<a/2\leq5\sqrt\epsilon$.


But then by the quantum Pinsker inequality we have that at the final time $T$: $||\rho-\sigma_T||_t\leq\sqrt{10\ln 2\sqrt\epsilon}<0.1$ in the end, and hence for {\em all} measurements their results are close. Hence no more than $O(q/\sqrt\epsilon)=O(q)$ messages have to be sent.

We now finish the proof by showing the claim.

Define another state $$\tilde{\sigma_i}=
\left(
\begin{array}{cc}
A\hspace{0cm}& 0\\
0& \hspace{0cm}D\\
\end{array}
\right),
$$

i.e., the state $\sigma_i$ with its upper right and lower left quadrants deleted. It is easy to see that this is still a density matrix. Indeed $\tilde{\sigma_i}$ is the state $\sigma_i$ after measuring $P_y,I-P_y$.
Also define $\tilde{\rho}$ to be the matrix $\rho$ with its upper left and lower right quadrants replaced by 0, which is again a density matrix resulting from measuring $\rho$.
If $$
\rho=\left(
\begin{array}{rr}
E &F\\
G &H\\
\end{array}
\right) \mbox{  then }
\tilde{\rho}=\left(
\begin{array}{rr}
E &0\\
0 &H\\
\end{array}
\right).$$

By Uhlmann monotonicity we have $S(\rho||\sigma_i)\geq S(\tilde{\rho}||\tilde{\sigma_i})$, and it suffices to show that
$S(\tilde{\rho}||\tilde{\sigma_i})\geq S(\rho||\sigma_{i+1})+a/2$. Note that both $\tilde{\sigma_i}$ and $\sigma_{i+1}$ are diagonal in the basis $B_i$. Furthermore $S(\rho||\sigma_{i+1})=S(\tilde{\rho}||\sigma_{i+1})+S(\tilde{\rho})-S(\rho)$.

We first bound the term $S(\tilde{\rho})-S(\rho)$. In the basis $B_i$ we may view $\rho$ as a bipartite state $\rho_{RQ}$ consisting of a qubit $R$ (corresponding to membership in $V_y$) and the remaining qubits $Q$. In $\tilde{\rho}$ the qubit $R$ has been measured. Consider attaching another qubit $T$, and instead of measuring $R$ applying the unitary that ``copies'' $R$ to $T$. After the unitary $S(\rho)=S(\rho_{QRT})\geq S(\rho_{QR})-S(\rho_T)=S(\tilde{\rho})-S(\rho_T)$, due to the Araki-Lieb inequality and because $\rho_{QR}=\tilde{\rho}$. But $S(\rho_T)=H(\epsilon_y)$ and hence  $S(\tilde{\rho})-S(\rho)\leq H(\epsilon_y)$.

We need to compare $\Tr(\tilde{\rho}\log\tilde{\sigma_i})$ and $\Tr(\tilde{\rho}\log\sigma_{i+1})$. Note that $\tilde{\sigma_i}$ and $\sigma_{i+1}$ are both diagonal.

\begin{eqnarray*}
&&\Tr(\tilde{\rho}\log\tilde{\sigma_{i}})\\
&=&\sum_j \tilde{\rho}(j,j) \log(\tilde{\sigma_i}(j,j))\\
&=&\sum_{j\leq d/2} \tilde{\rho}(j,j)\log(\sigma_{i+1}(j,j)\cdot \frac{1-a}{1-\epsilon_y})\\
&+&\sum_{j> d/2} \tilde{\rho}(j,j)\log(\sigma_{i+1}(j,j)\cdot \frac{a}{\epsilon_y})\\
&=&\sum_j \tilde{\rho}(j,j)\log(\sigma_{i+1}(j,j))+(1-\epsilon_y)\cdot\log\frac{1-a}{1-\epsilon_y}+\epsilon_y\cdot\log\frac{a}{\epsilon_y}\\
&=&\Tr(\tilde{\rho}\log\sigma_{i+1})+H(\epsilon_y)-H(\epsilon_y,a),
\end{eqnarray*}
where $H(u,v)=-u\log v-(1-u)\log(1-v)$.

Assuming that $a\geq10\sqrt\epsilon$ and $\epsilon_y\leq\epsilon=10^{-6}$ we can estimate
\begin{equation}\label{eq:H} H(\epsilon_y,a)\geq a.\end{equation}
If $a>1/2$ then (\ref{eq:H}) is true from the first term $(1-\epsilon_y)\log(1/(1-a))\geq a$.
Otherwise for all $a\in[0,1]$ we have $-\log (1-a)\leq a$, hence the first term in $H(\epsilon_y,a)$ is at least $(1-\epsilon_y)a$. And for the second term $\epsilon_y\log (1/a)\geq\epsilon_y\log(2)\geq\epsilon_y$, so (\ref{eq:H}) is always true.

Also note that $H(\epsilon_y)\leq H(\epsilon)\leq \epsilon\log(\epsilon^{-1})\leq\sqrt\epsilon\leq a/10$.

But then
\begin{eqnarray*}
&&S(\rho||\sigma_i)\\
&\geq &S(\tilde{\rho}||\tilde{\sigma_i})\\
&=&-S(\tilde{\rho})-\Tr(\tilde{\rho}\log\tilde{\sigma_i})\\
&\geq&-S(\rho)-H(\epsilon)-\Tr(\tilde{\rho}\log\tilde{\sigma_i})\\
&\geq & -S(\rho)-H(\epsilon)-\Tr(\tilde{\rho}\log\sigma_{i+1})+a-H(\epsilon)\\
&=&-S(\rho)-\Tr(\rho\log\sigma_{i+1})+a-2H(\epsilon)\\
&\geq&S(\rho||\sigma_{i+1})+a/2.\end{eqnarray*}

\end{proof}

\section{Quantum versus Classical Proofs}

Let us first define the problem for which we prove our separation result.

 \begin{definition}
   The function $\MajIx(x,I)$, where $I=\{i_1,\ldots, i_{\sqrt n}\}$, each $i_j\in\{1,\ldots,n\}$, and $x\in\{0,1\}^n$ is defined as follows:
   \begin{enumerate}
     \item if $|\{j:x_{i_j}=1\}|=\sqrt n$ then $\MajIx(x,I)=1$,
     \item if $|\{j:x_{i_j}=1\}|\leq 0.9\sqrt n$ then $\MajIx(x,I)=0$,
    \item  otherwise $\MajIx(x,I)$ is undefined.
   \end{enumerate}
 \end{definition}

The function has been studied in \cite{klauck:qma}, where it is shown that one-way MA protocols for the problem need communication $\Omega(\sqrt n)$. Our main technical result here is to extend this to one-way QCMA protocols.

It is obvious on the other hand, that there is a cheap protocol when Bob can send a message to Alice.

\begin{lemma}\label{lem:qma}
$R^{B\to A}(\MajIx)=O(\log n)$.\end{lemma}

Raz and Shpilka \cite{raz:qproof} show that any problem with $QMA(f)=c$ (i.e., QMA protocol where Alice and Bob can interact over many rounds) has a QMA protocol of cost $poly(c)$ in which Merlin sends a message to Alice, who sends a message to Bob. By inspection of their proof the polynomial overhead can be removed in the case of constant rounds of interaction between Alice and Bob.

\begin{lemma}\label{lem:upper}
If $QMA(f)=c$ and this cost can be achieved by a protocol with $O(1)$ rounds, then $QMA^{A\to B}(f)=O(c)$.
\end{lemma}

We give more details in Appendix~\ref{app:upperlem}. The lemma immediately implies the following.
\begin{theorem}\label{thm:upper}
$QMA^{A\to B}(\MajIx)=O(\log n)$.
\end{theorem}

We give a self-contained proof of this fact in Appendix~\ref{app:upperthm}. Our protocol has completeness 1, hence even the one-sided error version of $QMA^{A\to B}$ is separated from $QCMA^{A\to B}$ by the following lower bound.

\begin{theorem}$QCMA^{A\to B,*}(\MajIx)\geq\Omega(\sqrt{n}/\log n)$.
\end{theorem}

Hence we can conclude the following.
\begin{corollary}
There is a partial Boolean function $f$ such that $QMA^{A\to B}(f)=O(\log n)$, while $QCMA^{A\to B,*}(f)=\Omega(\sqrt n/\log n)$.
\end{corollary}

\begin{proof}
Fix any QCMA protocol $\cal P$ for $\MajIx$. Furthermore define a distribution on inputs as follows.
Fix any error correcting code $C\subseteq \{0,1\}^n$ with distance $n/4$ (i.e., every two codewords have Hamming distance at least $n/4$). Such codes of size $2^{\Omega(n)}$ exist by the Gilbert-Varshamov bound. We do not care about the complexity of decoding and encoding for our code. Furthermore we require the code to be balanced, i.e., that any codeword has exactly $n/2$ ones. This can also be achieved within the stated size bound.
For our distribution on inputs first choose $x\in C$ uniformly, and then uniformly choose $I$ among all subsets of $\{1,\ldots,n\}$ of size $\sqrt n$. Note that the probability of 1-inputs
under the distribution $\mu$ just defined is between $2^{-\sqrt n}$ and $2^{-\sqrt n-1}$ due to the balance condition on the code.

If the cost (i.e., communication from Merlin plus communication from Alice) of $\cal P$ is $c$, then there are at most $2^c$ different classical proofs sent by Merlin. We identify such proofs $p$ with the set of 1-inputs that are accepted by the protocol with probability at least $2/3$ when using the proof $p$. Hence there must be a proof $P$ containing 1-inputs of measure at least $2^{-\sqrt  n-c-1}$, because for every 1-input there is a proof with which it is accepted with probability $2/3$ or more.
Furthermore, given $P$, no 0-input is accepted with probability larger than $1/3$. Note that inputs outside of the promise, or 1-inputs outside of $P$ can be accepted with any probability between 0 and 1.
Denote by $f_P$ the partial function $\{0,1\}^n\times\{0,1\}^m\to\{0,1,\bot\}$, in which all inputs in $P$ are accepted, and all 0-inputs of $f$ are rejected, and the remaining inputs have undefined function value ($\bot$). $m=\Theta(\sqrt n\log n)$ is the length of Bob's input. Denote by $M={n\choose \sqrt n}$ the  number of Bob's inputs.

Obviously $f_P$ can be computed  by a one-way quantum protocol without prover using communication $c$ (and possibly using shared entanglement between Alice and Bob). Now due to Theorem \ref{thm1} this implies that $D^{A\to B}(f_P)\leq O(c\cdot m)$. We will argue that on the other hand $D^{A\to B}(f_P)\geq\Omega(n)$, and hence $c\geq \Omega(n/m)=\Omega(\sqrt{n}/\log n)$, which is our theorem.

Denote by $A$ the communication matrix of $f_P$. A row of $A$ is {\em fat}, if it contains more than $M2^{-\sqrt n-2c}$ 1-inputs (to $f_P$). Note that there can be no fewer than $|C|2^{-c-2}$ fat rows, because there are at least $|C|M2^{-\sqrt n-c-1}$ 1-inputs in $P$, the non-fat rows contain together at most $|C|M2^{-\sqrt n-2c}$ 1-inputs and each fat row at most $M2^{-\sqrt n}$ 1-inputs. Let $C'$ denote the row set consisting of the fat rows only, and $A'$ the matrix $A$ restricted to those rows. We claim that $A'$ has $C'$ distinct rows. Recall that for distinct rows there is a column, where one row has a 1 entry and the other a 0 entry. This means that $D^{A\to B}(f_P)\geq\log C'\geq \log |C|-c-2\geq\Omega(n)$ (unless $c=\Omega(n)$ already).

To show that all pairs of rows in $A'$ are distinct consider two of them, named $x,y$. We identify the row labeled by $x$ with the set for which $x$ is the characteristic vector. Recall $x,y$ are both codewords and are both fat. $x\cap y\leq n(1/2-1/8)$ because $x$ and $y$ have Hamming distance at least $n/4$. Let $S\subseteq\cal I$ be the set of $I$ such that $(x,I)\in P$. Then for all $I\in S$ we have that all of the $i\in I$ must satisfy $x_i=1$.

Furthermore let $T\subseteq S$ be the set $I\in S$, such that $|I\cap x\cap y|\geq 0.9\sqrt n$. Then
\[|T|\leq \sqrt n\cdot{\frac{3n}{8}\choose {0.9\sqrt n}}\cdot {\frac{n}{8}\choose {0.1\sqrt n}}   \leq 2^{-\alpha \sqrt n}\cdot{\frac{n}{2}\choose {\sqrt n}}   ,  \]
for some  constant $\alpha>0$. Note that the binomial coefficient on the right hand side is the number of $I\in\cal I$ such that $(x,I)$ is a 1-input. Hence
\[\frac{\mu((\{x\}\times T)\cap P)}{\sum_{I\in {\cal I}: \MajIx(x,I)=1} \mu(x, I)}\leq 2^{-\alpha \sqrt n},\]
i.e., a small fraction of 1-inputs $(x,I)$ in $P$ on row $x$ have $I\in T$, but $x$ is fat and has more 1-inputs.
 We can assume that $c<\alpha\sqrt n/10$ and hence $2^{-2c}\geq 2\cdot2^{-\alpha\sqrt n}$, so that the set $T$ contributes little to the set of $I\in \cal I$ with  $(x,I)\in P$. In particular there must be at least one $I\not\in T$ such that $(x,I)$ is in $P$, and hence $f_P(x,I)=1$.

So let us examine the set  of all $I\in S-T$ (so $|I\cap x\cap y|< 0.9\sqrt n$). $|I\cap x|=\sqrt n$ and hence $|y\cap I|\leq .9\sqrt n$, i.e., $(y,I)$ is a 0-input. This gives us the desired column $I$, such that $f_P(x,I)=1$ and $f_P(y,I)=0$, i.e., $x,y$ are two distinct rows in $A'$.

\end{proof}

\sect[concl]{Open Problems}

\begin{itemize}
\item Aaronson \cite{aaronson:qlearnability} argues that the bound in Theorem \ref{thm1} is tight to within polylogarithmic factors for partial functions. However, a longstanding conjecture is that for all total Boolean functions $f$ we have $R^{A\to B}(f)=O(Q^{A\to B}(f))$, or something weaker, i.e., that for total Boolean functions we can remove the factor $m$ completely (possibly at the expense of increasing the dependence on $Q^{A\to B}(f)$ polynomially). An easier problem might be to replace $m$ by something smaller like $\sqrt m$ for total functions.
\item For many "nondeterministic" modes of communication complexity one-way communication is as good as two-way communication, for instance for nondeterministic, QMA, AM-complexity. We have proved that this is not the case for QCMA protocols (and this was known previously for MA-protocols \cite{klauck:qma}). Is there a proper round-hierarchy for QCMA or MA protocols, i.e, is it true that there is a function that can be computed efficiently in $k$ rounds but not in $k-1$ rounds?
\item MA-communication complexity has recently been applied to the analysis of cloud-computing on data-streams and related topics \cite{Chakra:AM,Chakrabarti09annotations}. Currently we don't have lower bounds larger than $\sqrt n$ for MA-communication complexity of explicit functions, while counting arguments show that most functions have complexity $\Omega(n)$. This gap is quite significant in practice. Can larger bounds be shown for an explicit function, at least in the one-way model, or the even more restricted {\em online} one-way model?
\item Lower bounds for the AM-communication complexity of any explicit function remain elusive.
\item It would be interesting to separate QCMA- and QMA- communication complexity in the general two-way communication model. Such a separation could be used in the algebrization framework \cite{AaronsonW09} to argue that showing QCMA=QMA (in the Turing machine world) would require nonalgebrizing techniques, and might also be thought of as evidence that these classes are not equal after all.
\item A similar problem, and maybe less ambitious, is to show that the QCMA- and QMA- query complexities of a Boolean problem are very different. However, while for communication complexity we have a good candidate for such a separation (the QMA-complete problem), we are not aware of a good candidate for the query complexity setting. Is there a complete (promise) problem for QMA-query complexity?

    \end{itemize}

\section*{Acknowledgement}
The authors thank Dmitry Gavinsky for helpful discussions.

\bibliographystyle{alpha}
\bibliography{qc}

\begin{appendix}

\section{Proof Idea of Lemma \ref{lem:upper}}\label{app:upperlem}

 The proof of Raz and Shpilka in \cite{raz:qproof} proceeds by showing that a problem LSD is complete for QMA communication complexity. It is easy to see that LSD can be computed by a QMA one-way protocol with logarithmic communication. The main difficulty is showing that any problem $f$ with $QMA(f)=C$ can be reduced to an instance of LSD of size $2^{poly(C)}$. Here we will argue that if the QMA complexity of $f$ is $C$ for protocols that have only $O(1)$ rounds between Alice and Bob, then the constructed LSD instance has size $2^{O(C)}$ only, and hence we get that $QMA^{A\to B}(f)=O(C)$.

 For the problem LSD (linear space distance), Alice and Bob each receive a subspace of $\mathbb{R}^n$ of dimension $n/4$. The distance between two subspaces $V,W$ is the minimum over all pairs of unit vectors from $V$ and $W$ of the euclidean distance between the vectors. For LSD the promise is that the spaces given to Alice and Bob are either very close (distance at most $0.1\sqrt 2$) or very far (distance at least $0.9\sqrt 2$). The 1-inputs of LSD are the pairs of spaces that are close.
 Note that there is a QMA one-way protocol for this problem of logarithmic cost: Merlin sends Alice a vector that is close to both subspaces as a quantum state, Alice measures the state with the observable consisting of her subspace and its complement, and if the state projects into her subspace she sends the projected state to Bob (otherwise she rejects), who measures with his observable. If both measurements succeed, they accept.

 The main lemma in the reduction of \cite{raz:qproof} is the following (their Lemma 19):
 \begin{lemma}
 If $f$ has a QMA protocol with proof length $p$, communication $c$, and $r$ rounds and error $1/r^4$, then $f$ can be reduced (by local operation by Alice and Bob) to an instance of LSD where the underlying space is $\mathbb{R}^{(r+1)2^{2(c+p)}}$, and the distance between the two spaces is at least $1/r^{1.5}$ in the case of 0-inputs, and at most $\sqrt 2/r^{2.5}$ in the case of 1-inputs.
 \end{lemma}

 This lemma is then combined with standard boosting ideas to improve the gap in the distance, but for us the fact that amplifying the success probability by parallel repetition in a QMA one-way protocol is possible suffices. Furthermore, in the  case of $r=O(1)$,
  the LSD instance provided by the above lemma already has a constant gap and the correct size of $2^{O(c+p)}$. Reducing a function $f$ to the LSD instance, and then running $O(1)$ times the protocol for LSD in parallel suffices to get a one-way QMA-protocol for $f$.

\section{Proof of Theorem \ref{thm:upper}}\label{app:upperthm}

We describe a protocol with constant gap, which can be amplified to have the usual soundness using standard techniques.

Alice holds a string $x\in\{0,1\}^n$, Bob a set $I$ of $\sqrt n$ distinct indices. Alice receives a proof from Merlin, which is supposed to be the uniform superposition $\sum_{i\in I}|i\rangle$. Alice attaches one more qubit, applies the unitary that maps $|i\rangle|a\rangle$ to $|i\rangle|x_i\oplus a\rangle$, then measures that extra qubit. If the result is 0 she rejects. Otherwise she discards the extra qubit and sends the remaining state to Bob. Bob measures this state with the observable that consists of the subspace spanned by the vector with 1 in all positions $i\in I$ and 0 elsewhere, and its orthogonal complement. He accepts, if this measurement projects onto the first subspace. Note that the communication is $\log n$ qubits from both Merlin and Alice.

If Merlin is honest (and $x_i=1$ for all $i\in I$), he sends the uniform superposition $\sum_{i\in I}|i\rangle$. Alice's measurement will not change this state (she effectively projects into the space spanned by the $|i\rangle$ with $x_i=1$). Bob's measurement
will accept with certainty. Hence the protocol has completeness 1.

Now assume that at most $0.9\sqrt n$ of the $i\in I$ satisfy that $x_i=1$. We may assume that Merlin's message is a pure state of the correct dimension. For each mixed state there is a pure state that will perform at least as well, and states with the wrong dimension will be rejected immediately. Again, unless Alice rejects, her measurement projects into the space spanned by $|i\rangle$ for which $x_i=1$. Denote this state by $|\psi\rangle=\sum_{i:x_i=1}\alpha_i|i\rangle$. Bob measures the observable consisting of $\mbox{span}(|\phi_I\rangle)$, where $|\phi_I\rangle=\sum_{i\in I}|i\rangle/n^{1/4}$, and its orthogonal complement. The probability of the measurement accepting is the squared inner product of $|\phi_I\rangle$ and $|\psi\rangle$. This value is $(\sum_{i\in I:x_i=1}\alpha_i/n^{1/4})^2\leq  (1/\sqrt n)(0.9\sqrt n)(\sum |\alpha_i|^2)\leq 0.9$.
Hence the protocol has soundness error 0.9, which can be improved by parallel repetition.

\end{appendix}

\end{document}

%% file: LocalHeader.tex
\documentclass[11pt, a4paper]{article}

\usepackage{ifthen}

\newcommand{\MyMacro}[4]{
\newboolean{#1}
\setboolean{#1}{#2}
\newcommand{#3}{\ifthenelse{\boolean{#1}}}
\newcommand{#4}{\ifthenelse{\not \boolean{#1}}}
}

\MyMacro{is_slides}{false}{\Ifslides}{\IfNotslides}

\MyMacro{is_poster}{false}{\Ifposter}{\IfNotposter}

\MyMacro{use_packages}{true}{\IfUsepackages}{\IfNotUsepackages}

\MyMacro{do_bib}{true}{\IfBib}{\IfNoBib}

\MyMacro{proofs_to_app}{false}{\IfProofsToApp}{\IfNotProofsToApp}

\MyMacro{is_llncs}{false}{\Ifllncs}{\IfNotllncs}

\MyMacro{is_conference}{false}{\IfConf}{\IfNotConf}

\newcommand{\ifndef}[2]{\ifthenelse{\isundefined{#1}}{#2}{}}

\newcommand{\mydef}[2]{\def#1{#2}}

\newcommand{\nospell}[1]{#1}   %

{}
\usepackage{amssymb}
\usepackage{amsmath}   %
{}\usepackage{amsthm}{}   %
\usepackage{latexsym}
\usepackage{amsfonts}
\usepackage{units}     %
\usepackage{psfrag}    %
\usepackage{url}    %
\usepackage{hyphenat}   %
{}
{}
\usepackage[dvips]{graphicx}
\usepackage[usenames,dvipsnames]{color}
{}
{}
{}

{}  %
{}  %
\ifndef{\theorem}{\newtheorem{theorem}{Theorem}[section]}
\ifndef{\lemma}{\newtheorem{lemma}[theorem]{Lemma}}
\ifndef{\corollary}{\newtheorem{corollary}[theorem]{Corollary}}
\ifndef{\conjecture}{\newtheorem{conjecture}[theorem]{Conjecture}}
\ifndef{\remark}{\theoremstyle{remark} \newtheorem{remark}{Remark}}
\ifndef{\proposition}{\newtheorem{proposition}{Proposition}}
\ifndef{\claim}{\newtheorem{claim}[theorem]{Claim}}
\ifndef{\result}{\newtheorem{result}{Result}}
\ifndef{\problem}{\newtheorem{problem}{Problem}}
\ifndef{\fact}{\newtheorem{fact}[theorem]{Fact}}
\newtheoremstyle{mydefinition}   %
{\topsep}{\topsep}   %
{\slshape}   %
{}   %
{\bfseries}   %
{.}   %
{ }   %
{}   %
{\theoremstyle{mydefinition}\newtheorem{definition}{Definition}}

\newtheoremstyle{myremark}   %
{\topsep}{\topsep}   %
{\slshape}   %
{}   %
{\bfseries\slshape}   %
{:}   %
{ }   %
{}   %
{\theoremstyle{myremark}}

\newtheoremstyle{myexample}   %
{\topsep}{\topsep}   %
{\itshape}   %
{}   %
{\slshape}   %
{:}   %
{ }   %
{\ul{\thmname{#1}}}   %
\ifndef{\example}{\theoremstyle{myexample} \newtheorem{example}{Example}}

\newtheoremstyle{myclaims}   %
{\topsep}{\topsep}   %
{\slshape}   %
{}   %
{\bfseries\itshape}   %
{}   %
{ }   %
{\thmname{#1}\thmnumber{ \!#2}.}   %
{\theoremstyle{myclaims}

\ifndef{\fact}{\newtheorem{fact}[theorem]{Fact}}
}
{}  %
{}  %

{}
\newtheoremstyle{anystatement}{\topsep}{\topsep}{\itshape}{}{\bfseries}{.}{ }{\anystatementname}
{\theoremstyle{anystatement}}
\newcommand{\anystatementname}{}

{}

\newcommand{\AuxNew}[4][]{#2{#3}[1][*]%
{\ifthenelse{\equal{*}{##1}} %
{\Ensuremath{#1{#4}}}%
{\ifthenelse{\equal{b}{##1}} %
{\Ensuremath{\mathbf{#4}}}%
{\ifthenelse{\equal{}{##1}} %
{\IfMathMode{#1{#4}}{#4}}{}}}}}

\newcommand{\newident}[3][*]{\ifthenelse{\equal{*}{#1}}%
{\AuxNew[\mathit]{\newcommand}{#2}{#3}} %
{\mydef{#2}{\Ensuremath{\mathit{#3}}}}} %

\newcommand{\newmat}[3][*]{\ifthenelse{\equal{*}{#1}}%
{\AuxNew{\newcommand}{#2}{#3}} %
{\mydef{#2}{\Ensuremath{#3}}}} %

\newcommand{\providemat}[3][*]{\ifthenelse{\equal{*}{#1}} %
{\AuxNew{\providecommand}{#2}{#3}} %
{\mydef{#2}{\Ensuremath{#3}}}} %

\newcommand{\providematarg}[2]{ %
\providecommand{#1}[1][]{\Ensuremath{#2}}}

\newcommand{\newfunction}[2]{ %
\newcommand{#1}[2][*]{\ifthenelse{\equal{*}{##1}}%
{\Ensuremath{#2{\left(##2\right)}}}%
{#2(##2)}}%
}

\newcommand{\MyMakeTheoMacros}[3]{
\newcommand{#2}[2][]{\ifthenelse{\equal{}{##1}}
{\begin{#1} ##2 \end{#1}}
{\begin{#1}\label{##1} ##2\end{#1}}}
\newcommand{#3}[3][]{\ifthenelse{\equal{}{##1}}
{\begin{#1}[##2] ##3 \end{#1}}
{\begin{#1}[##2]\label{##1} ##3\end{#1}}}
}

\newcounter{tmp_id_cnt}
\newcommand{\MyMakeDupTheoMacros}[7]{
\MyMakeTheoMacros{#1}{#2}{#3}
\newcommand{#4}[3]{
\newcommand{##2}{##3}
\begin{#1}\label{##1} ##2\end{#1}}
\newcommand{#5}[4]{
\newcommand{##2}{##4}
\begin{#1}{\e{##3}}\label{##1} ##2\end{#1}}
\newcommand{#7}[2]{\def\my_tmp_id{my_tmp_id_\arabic{tmp_id_cnt}}
\newtheorem*{\my_tmp_id}{#6~\ref{##1}}
\begin{\my_tmp_id} ##2 \end{\my_tmp_id}\stepcounter{tmp_id_cnt}}
}

\newcommand{\MyMakeRefMacros}[3]{\newcommand{#1}[2][]
{\ifthenelse{\equal{}{##1}}{#2~\ref{##2}}{#3~\ref{##1} and~\ref{##2}}}}

\newcommand{\MyMakeEqRefMacros}[3]{\newcommand{#1}[2][]
{\ifthenelse{\equal{}{##1}}{#2~\eqref{##2}}{#3~\eqref{##1} and~\eqref{##2}}}}

{}
{}   %
\newcommand{\bibentry}[8]{
{}\bibitem[\nospell{#8}]{#1} {\textup #3}.{}
\ifthenelse{\equal{}{#6}}
{\newblock \textrm{#4.} \newblock {\em #5}, #7.}
{\newblock \textrm{#4.} \newblock {\em #5, #6}, #7.}
}

{}  %
{}  %

\MyMakeTheoMacros{fact}{\fct}{\nfct}

\MyMakeRefMacros{\fctref}{Fact}{Facts}

{}  %
\MyMakeDupTheoMacros{lemma}
{\lem}{\nlem}{\lemdup}{\nlemdup}{Lemma}{\lemrep}
{}

\MyMakeRefMacros{\lemref}{Lemma}{Lemmas}

\MyMakeDupTheoMacros{corollary}
{\crl}{\ncrl}{\crldup}{\ncrldup}{Corollary}{\crlrep}

\MyMakeRefMacros{\crlref}{Corollary}{Corollaries}

\MyMakeTheoMacros{proposition}{\prp}{\nprp}

{}
\newtheorem*{prp*}{\e{Proposition}}

{}

\MyMakeRefMacros{\prpref}{Proposition}{Propositions}

\MyMakeDupTheoMacros{my_claim}
{\clm}{\nclm}{\clmdup}{\nclmdup}{Claim}{\clmrep}

\MyMakeRefMacros{\clmref}{Claim}{Claims}

\MyMakeDupTheoMacros{theorem}
{\theo}{\ntheo}{\theodup}{\ntheodup}{Theorem}{\theorep}

\MyMakeRefMacros{\theoref}{Theorem}{Theorems}

\MyMakeTheoMacros{definition}{\defi}{\ndefi}

\MyMakeRefMacros{\defiref}{Definition}{Definitions}

\MyMakeTheoMacros{problem}{\prob}{\nprob}

\MyMakeRefMacros{\probref}{Problem}{Problems}

\MyMakeTheoMacros{conjecture}{\conj}{\nconj}

\MyMakeRefMacros{\conjref}{Conjecture}{Conjectures}

\providecommand{\qedsymbol}{\square}

\newcommand{\prf}[2][]{\ifthenelse{\equal{}{#1}}%
{\begin{proof}\renewcommand{\qedsymbol}{$\blacksquare$}%
#2 \end{proof}}%
{\begin{proof}[Proof of #1]%
\renewcommand{\qedsymbol}{$\blacksquare_{\mbox{\it{\scriptsize{#1}}}}$}%
#2 \end{proof}}
}

\newcommand{\abstr}[1]{\begin{abstract} #1 \end{abstract}}

\newcommand{\sect}[2][]{\ifthenelse{\equal{}{#1}}
{\section{#2}}
{\section{#2}\label{#1}}}

\MyMakeRefMacros{\chref}{Chapter}{Chapters}

\MyMakeRefMacros{\sref}{Section}{Sections}

\MyMakeRefMacros{\ssref}{Subsection}{Subsections}

\MyMakeRefMacros{\sssref}{Subsection}{Subsections}

\MyMakeRefMacros{\figref}{Figure}{Figures}

\newcommand{\IfMathMode}[2]{\ifmmode{#1}\else{#2}\fi}

\newcommand{\Ensuremath}{\ensuremath}

\newcommand{\fbr}[1]{\IfMathMode %
{#1}{$#1$}}                      %

\newcommand{\fnbr}[1]{\mbox{\fbr{#1}}}   %

\newcommand{\fla}[2][*]{\ifthenelse{\equal{}{#1}}{\fbr{#2}}{\fnbr{#2}}}

\newcommand{\mat}[2][]{\ifthenelse{\equal{}{#1}} %
{ \begin{displaymath} #2 \end{displaymath} } %
{ \begin{equation} \label{#1} #2 \end{equation} }%
}

\MyMakeEqRefMacros{\equref}{Equation}{Equations}

\MyMakeEqRefMacros{\expref}{Expression}{Expressions}

\MyMakeEqRefMacros{\inequref}{Inequality}{Inequalities}

\providecommand{\middle}{\big}

\newcommand{\h}[2][]{\ifthenelse{\equal{}{#2}}%
{h_{#1}}%
{h_{#1}{\left[{#2}\right]}}}

\newcommand{\I}[3][]{\ifthenelse{\equal{}{#1}}%
{\mathbf{I}{\left[{#2}:{#3}\right]}}%
{\mathbf{I}{\left[{#2}:{#3}\middle|{#1}\right]}}}

\providecommand{\E}[2][]{\ifthenelse{\equal{}{#1}}%
{\mathop{\mathbf{E}}{\left[{#2}\right]}}%
{\mathop{\mathbf{E}}_{#1}{\left[{#2}\right]}}}

\newcommand{\ord}[1][]{\nospell{\ifthenelse{\equal{}{#1}}%
{\txt{'th}}%
{\ifthenelse{\equal{1}{#1}}{$1\txt{'st}$}{\ifthenelse{\equal{2}{#1}}{$2\txt{'nd}$}{\ifthenelse{\equal{3}{#1}}{$3\txt{'rd}$}{\fla{#1\txt{'th}}}}}}}}

\newcommand{\fr}[3][*]{%
\ifthenelse{\equal{*}{#1}}        %
{\frac{#2}{#3}}{}%
\ifthenelse{\equal{/}{#1}}        %
{\nicefrac{#2}{#3}}{}%
\ifthenelse{\equal{}{#1}}         %
{\left.#2\middle/#3\right.}{}%
\ifthenelse{\equal{p_}{#1}}       %
{\left.\left(#2\right)\middle/#3\right.}{}%
\ifthenelse{\equal{_p}{#1}}       %
{\left.#2\middle/\left(#3\right)\right.}{}%
\ifthenelse{\equal{pp}{#1}}       %
{\left.\left(#2\right)\middle/\left(#3\right)\right.}{}
}

\def\MySQRT#1#2{    %
\setbox0=\hbox{$#1\sqrt{#2\,}$}\dimen0=\ht0%
\advance\dimen0-0.2\ht0%
\setbox2=\hbox{\vrule height\ht0 depth -\dimen0}%
{\box0\lower0.4pt\box2}}

\newcommand{\set}[2][]{\ifthenelse{\equal{}{#1}} %
{\Ensuremath{\left\{#2\right\}}}%
{\Ensuremath{\left\{#2\,\middle\arrowvert\,#1\right\}}}}

\newfunction{\asO}{O}
\newfunction{\aso}{o}
\newfunction{\asOm}{\Omega}
\newfunction{\astOm}{\tilde \Omega}

\mydef{\01}{\set{0,1}}
\mydef{\12}{\set{1,2}}

\newcommand{\txt}[1]{\textrm{#1}}   %

\DeclareMathAlphabet{\lowcal}{OT1}{pzc}{m}{it}

\newmat[]{\llp}{\left(}
\newmat[]{\rrp}{\right)}

\newmat[]{\dt}{\cdot}
\newmat[]{\tm}{\cdot}
\newmat[]{\deq}{\stackrel{\textrm{def}}{=}}

\providemat{\QQ}{\mathbb{Q}}
\providematarg{\NN}{\ifthenelse{\equal{}{#1}}%
{\mathbb{N}}%
{\mathbb{N}_{#1}}}

\newcommand{\ds}[1][]
{\ifthenelse{\equal{}{#1}}{\dots}{#1\dots#1}}
\newmat[]{\dc}{\ds[,]}

\newcommand{\itemi}[2][]{\ifthenelse{\equal{}{#1}}
{\begin{itemize} #2 \end{itemize}}
{\begin{itemize}[#1] #2 \end{itemize}}}

{}  %
\newcommand{\e}{\emph}
{}   %

\providecommand{\ul}[1]{\underline{#1}}  %

{}
{}